\documentclass[emptyheadfoot]{andp2012}
\usepackage{graphicx}
\usepackage[english]{babel}
\setcopyrightyear{2015}%
\setyear{2015}%
\DOIprefix{10.1002}%
\DOIsuffix{andp.201500259}
\Volume{527}%
\Issue{11-12}%
\Month{11}%
\Year{2015}%
\pagespan{794}{805}%
\category{Original paper}%
\shortabstract%
\begin{abstract}
Ground-state properties of the non-interacting symmetric 
single-impurity Anderson model (SIAM)
are derived from the corresponding eigenenergy equation.
Explicit formulae are given for the ground-state energy,
the hybridization, and the momentum distribution that are essential quantities
for variational approaches to the interacting model. 
Various spectral functions, e.g.,
the total density of states, the phase shift function, 
and the impurity spectral function,
are shown to agree with those obtained from the equation-of-motion method
(see supplementary material).
For a constant hybridization strength and a semi-elliptic host density of states
it is seen that the impurity spectral function builds up weight at the band 
edges.
\end{abstract}

\title[Non-interacting single-impurity Anderson model]%
{\centerline{Non-interacting single-impurity Anderson model:}
\centerline{solution without using the equation-of-motion method}}
\author{Zakaria M.M.\ Mahmoud\inst{1,2}}
\author{Florian Gebhard\inst{2}\fnmsep\footnote{Corresponding author\quad%
E-mail: florian.gebhard@physik.uni-marburg.de}}
\shortauthors{Z.\, M.\, M.\ Mahmoud and F.\ Gebhard}
\address[\inst{1}]{Department of Physics, New-Valley Faculty of Science, 
El-Kharga, Assiut University, Egypt}
\address[\inst{2}]{Fachbereich Physik, Philipps-Universit\"at Marburg,
D-35032 Marburg, Germany}

\begin{document}

\maketitle

\section{Introduction}
\label{sec:intro}
The single-impurity Anderson model~\cite{Anderson1961}
is one of the fundamental many-body
problems in condensed-matter theory. Even its non-interacting limit 
poses a non-trivial single-particle problem because 
the electrons on a single site hybridize with those from a conduction band
with a large (or infinite) number of degrees of freedom.

The non-interacting single-impurity Anderson model 
can be solved exactly~\cite{Anderson1961,Hewsonbook}.
Usually, we are interested in 
the ground-state energy, the density of states,
and single-particle Green functions at zero temperature.
In textbooks~\cite{Hewsonbook,Solyombook},
these single-particle properties 
are calculated from the equation-of-motion approach for the
single-particle Green functions. 
In this communication, we derive them directly from 
the exact eigenstates and eigenenergies.
Apart from being instructive for beginners in many-body theory,
the direct approach facilitates a comparison with numerical approaches 
and covers all general cases (finite bandwidth of the conduction band, 
bound and anti-bound states).
Moreover, ground-state expectation values are important for 
variational approaches such as the Gutzwiller wave function,
see, e.g.,~\cite{Schoenhammer1990,Gebhard1991},
so that it is important to have general expressions available
for the hybridization and momentum distribution functions.
To simplify the discussion, we focus on the symmetric single-impurity Anderson model.

Our work is organized as follows. In chapter~\ref{sec:model}, 
we introduce the model Hamiltonian. In chapter~\ref{sec:GreenF}
we discuss the single-particle Green functions and some of their properties. 
In chapter~\ref{sec:eigen},
we solve the Schr\"odinger equation where we treat bound/anti-bound states
and scattering states separately; we shall take the thermodynamic limit
where appropriate. In chapter~\ref{sec:gsexpectationvalues} we derive
a variety of ground-state quantities, namely  the total energy, 
the impurity occupancy, 
the hybridization energy, and the momentum distribution.
In chapter~\ref{sec:eigenspec} we consider the
single-particle spectral properties and derive the density of states, 
the phase shift function, and the impurity spectral function.
For comparison,
in the supplementary material we
re-derive all expressions from the standard 
equation-of-motion approach.
Short conclusions, chapter~\ref{sec:conclusions}, close our presentation.

\section{Model and physical quantities}
\label{sec:model}

The Hamiltonian of the single-impurity Anderson model consists
of three parts, the host kinetic energy $\hat{T}$,
the impurity level $\hat{I}$, and the hybridization $\hat{V}$,
\begin{equation}
\hat{H}= \hat{H}_0+ \hat{I} \quad , \quad
\hat{H}_0=\hat{T} +\hat{V} \;.
\end{equation}
The eigenstates of the model are denoted by $|\psi_n\rangle$.
Their energy is $E_n$, 
\begin{equation}
\hat{H}|\psi_n\rangle = E_n |\psi_n\rangle \; ,
\label{eq:setofeigenstates}
\end{equation}
and $|\psi_0\rangle$ is the ground state with energy $E_0$.

\subsection{Host electrons}

We consider a host system of non-interacting electrons that is described
by the host density of states $\rho_0(\epsilon)$ in the thermodynamic limit.
Since the thermodynamic limit can be delicate for a single impurity in a bath,
we discretize the bath levels and describe the host electrons by their
kinetic energy
\begin{equation}
\hat{T}= \sum_{k,\sigma} \epsilon(k) 
\hat{c}_{k,\sigma}^+\hat{c}_{k,\sigma}^{\vphantom{+}} \; ,
\end{equation}
where $\hat{c}_{k,\sigma}^+$ ($\hat{c}_{k,\sigma}^{\vphantom{+}}$) creates (annihilates)
a spin-$\sigma$ electron in the quantum state~$k$ 
($\sigma=\uparrow,\downarrow$).
For a finite system with $L$ states we have $k=0,1,2,\ldots, (L-1)$,
and we choose $L$ to be an odd number for convenience.
The thermodynamic limit corresponds to $L\to\infty$.

The band energies are given by the dispersion relation
\begin{equation}
\epsilon(k) = f(-W/2+k W/(L-1))  \;,
\label{eq:defineepsasffunction}
\end{equation}
where $f(-\varepsilon)=-f(\varepsilon)$ 
is an odd, differentiable and mono\-tonously increasing 
function that defines the symmetric 
host density of states $\rho_0(\epsilon)= \rho_0(-\epsilon)$,
\begin{equation}
\rho_0(\epsilon) 
= \frac{1}{f'\left(f^{-1}(\epsilon)\right) } \; .
\label{eq:defrhozero}
\end{equation}
Later, we shall work with dispersion relations that satisfy
\begin{equation}
f(\pm W/2)=\pm W/2
\end{equation}
so that $W>0$ defines the host electron bandwidth. 
In the following we shall use $W\equiv 1$ as our energy unit.
Note that $f(0)=0$, i.e., there is a host state at $k=(L-1)/2$ with 
zero kinetic energy, $\epsilon((L-1)/2)=0$, and that
the completely filled host band has total energy zero,
$\sum_k\epsilon(k)=0$ because $\epsilon(L-1-k)=-\epsilon(k)$.

Later, we shall give explicit results for a semi-elliptic density of states,
\begin{equation}
\rho_0^{\rm se}(\epsilon) = \frac{4}{\pi} \sqrt{1-\left(2\epsilon\right)^2\, }
\quad , \quad 
|\epsilon| \leq 1/2
\label{eq:rhosemiellipse}
\end{equation}
with
\begin{equation}
\int_{-1/2}^{1/2} {\rm d}\epsilon 
\rho_0^{\rm se}(\epsilon) = 1  \; .
\end{equation}
The function $f_{\rm se}(\varepsilon)$ 
that leads to the semi-elliptic density of states~(\ref{eq:rhosemiellipse})
solves the implicit equation
\begin{equation}
\pi \varepsilon = 2 f_{\rm se}(\varepsilon)
 \sqrt{1-4 [f_{\rm se}(\varepsilon)]^2}+\sin^{-1}[2f_{\rm se}(\varepsilon)] \; ,
\end{equation}
where $\sin^{-1}(z)={\rm arcsin}(z)$ is the inverse sine function.
In some cases we shall also give the result for a constant density of states,
$f_{\rm cons}(\varepsilon)=\varepsilon$, 
$\rho_0^{\rm cons}(\epsilon)=1$ for $|\epsilon|\leq 1/2$.
Note, however, 
that the constant density of states has some pathological features,
e.g., a jump discontinuity at the band edges.

\subsection{Impurity level}

The impurity level is described by the Hamiltonian
\begin{equation}
\hat{I} = E_{\rm d} \sum_{\sigma} \hat{d}_{\sigma}^+\hat{d}_{\sigma}^{\vphantom{+}}
+ U \hat{d}_{\uparrow}^+\hat{d}_{\uparrow}^{\vphantom{+}}
\hat{d}_{\downarrow}^+\hat{d}_{\downarrow}^{\vphantom{+}}
\; .
\end{equation}
Here, $E_{\rm d}$ is the energy of the impurity level, and $U$ is the $d$-electrons'
Hubbard interaction. 
For the symmetric single-im\-purity Anderson model, we place the
impurity level at the particle-hole symmetric energy
\begin{equation}
E_{\rm d}=-U/2 \; .
\end{equation}
Later  we only address the non-interacting case, $U=0$.

\subsection{Hybridization}

The host electrons and the impurity level can hybridize via
\begin{equation}
\hat{V} = \sqrt{\frac{1}{L}} \sum_{k,\sigma}
\left( 
V_{k,\sigma}\hat{d}_{\sigma}^+\hat{c}_{k,\sigma}^{\vphantom{+}}
+
V_{k,\sigma}^*\hat{c}_{k,\sigma}^+\hat{d}_{\sigma}^{\vphantom{+}}
\right) \; ,
\end{equation}
where the amplitude $V_{k,\sigma}$ parameterizes the hybridization
strength. We demand that the hybridization is independent of spin,
$V_{k,\sigma}\equiv V_k$.
Since $\epsilon(k)$ is a monotonous function
of~$k$, we may equally write 
\begin{equation}
V_k\equiv V(\epsilon(k))\; .
\end{equation}
Furthermore, we demand that $V_k$ is symmetric,
$V_k=V_{L-1-k}^*$, in order to ensure
particle-hole symmetry.

\subsection{Particle-hole symmetry}

We study the case of half band-filling where 
the total number of electrons $N=N_{\uparrow}+N_{\downarrow}$
equals the total number of levels in the system, $N=L+1$.
For the paramagnetic case of interest, 
we then have $N_{\uparrow}=N_{\downarrow}=N/2=(L+1)/2$.

The particle-hole transformation is defined by
\begin{eqnarray}
\widetilde{\tau}: 
& \hat{c}_{k,\sigma}^{\vphantom{+}} \mapsto \hat{c}_{L-1-k,\sigma}^+ \quad ,\quad
 \hat{c}_{k,\sigma}^+ \mapsto \hat{c}_{L-1-k,\sigma}^{\vphantom{+}} \; ,\nonumber \\
& \hat{d}_{\sigma}^{\vphantom{+}} \mapsto -\hat{d}_{\sigma}^+
\quad , \quad \hat{d}_{\sigma}^+ \mapsto -\hat{d}_{\sigma}^{\vphantom{+}} \; .
\end{eqnarray}
The transformation leaves the Hamiltonian invariant, i.e.,
$\hat{H}\mapsto\hat{H}$, because we have $\epsilon(L-1-k)=-\epsilon(k)$
and $V_k=V_{L-1-k}^*$. Consequently, the same applies to the ground state,
$| \psi_0\rangle \stackrel{\tilde{\tau}}{\mapsto} | \psi_0\rangle$. 
Therefore, we can derive
the following relations at half band-filling
for the ground-state expectation values
\begin{eqnarray}
n_{k,\sigma} &=& \langle 
\hat{c}_{k,\sigma}^+\hat{c}_{k,\sigma}^{\vphantom{+}} 
 \rangle 
= 1 - n_{L-1-k,\sigma} \; , \label{eq:nkphsymm}\\
\langle 
\hat{d}_{\sigma}^+\hat{c}_{k,\sigma}^{\vphantom{+}} 
 \rangle 
&=&
\langle 
\hat{c}_{L-1-k,\sigma}^+\hat{d}_{\sigma}^{\vphantom{+}} 
 \rangle \; , \\
n_{{\rm d },\sigma} &=& \langle 
\hat{d}_{\sigma}^+\hat{d}_{\sigma}^{\vphantom{+}} 
 \rangle 
= 1- n_{{\rm d},\sigma} = \frac{1}{2} \; ,
\label{eq:dishalf}
\end{eqnarray}
where
\begin{equation}
\langle \hat{A}\rangle = \frac{
\langle \psi_0 | \hat{A} | \psi_0\rangle 
}{\langle \psi_0 | \psi_0\rangle } \; .
\end{equation}
Equation~(\ref{eq:dishalf}) proves that the $d$-level is half filled
for any dispersion relation and hybridization, $n_{{\rm d},\sigma}=1/2$.
Note that the relations~(\ref{eq:nkphsymm})--(\ref{eq:dishalf}) 
apply to the interacting case, $U\geq 0$.

\section{Single-particle Green functions}
\label{sec:GreenF}

\subsection{Retarded, advanced, and causal Green functions}

For Heisenberg operators ($\hbar\equiv 1$)
\begin{equation}
\hat{A}(t) = e^{{\rm i}\hat{H} t}\hat{A} e^{-{\rm i}\hat{H} t} 
\end{equation}
we consider the causal Green function
\begin{equation}
G_{A,B}^{\rm c}(t) = (-{\rm i}) \langle {\cal T} \left( 
\hat{A}(t)\hat{B}\right)\rangle \; ,
\end{equation}
where ${\cal T}$ is the time-ordering operator,
\begin{equation}
{\cal T}\left(\hat{A}(t)\hat{B}\right) = 
\left\{  
\begin{array}{@{}lcl@{}}
\hphantom{-}\hat{A}(t)\hat{B} & \hbox{for} & t>0 \\
-\hat{B}\hat{A}(t) & \hbox{for} & t<0 
\end{array}
\right. \; .
\end{equation}
The sign applies for Fermion operators $\hat{A},\hat{B}$.
The retarded and advanced Green functions are defined by
\begin{eqnarray}
G_{A,B}^{\rm ret}(t) &=& (-{\rm i}) \Theta(t) 
\langle  \left[ \hat{A}(t),\hat{B}\right]_+ \rangle \; , \nonumber \\
G_{A,B}^{\rm adv}(t) &=& {\rm i} \Theta(-t) 
\langle \left[ \hat{A}(t),\hat{B} \right]_+ \rangle \; ,
\label{eq:defGFretadvanced}
\end{eqnarray}
where $\Theta(t)$ is the Heaviside step-function.

\subsection{Fourier transformation}

For later use we introduce the Fourier transformation  (FT)
\begin{eqnarray}
f(t) &=& \int_{-\infty}^{\infty} \frac{{\rm d}\omega}{2\pi} 
e^{-\eta |\omega|} e^{-{\rm i}\omega t} 
\tilde{f}(\omega) \;, \nonumber\\
\tilde{f}(\omega) &=& \int_{-\infty}^{\infty} {\rm d}t e^{-\eta |t|}
e^{{\rm i}\omega t} f(t) \;, \label{eq:Fouriertransformation}
\end{eqnarray}
where the factors $\exp(-\eta |\omega|)$ and $\exp(-\eta |t|)$
with $\eta=0^+$ ensure the convergence of the integrals.
They shall be set to zero whenever the convergence of integrals 
or other expressions is guaranteed at $\eta=0$.

We use a complete set of eigenstates for the Hamiltonian $\hat{H}$,
see equation~(\ref{eq:setofeigenstates}),
to derive the Lehmann representation of the causal and retarded Green functions,
\begin{eqnarray}
 \tilde{G}_{A,B}^{\rm c}(\omega) &=&
\sum_n \biggl[ 
\frac{
\langle \psi_0| \hat{A}|\psi_n\rangle \langle \psi_n | \hat{B} | \psi_0\rangle}{
E_0-E_n+\omega+{\rm i}\eta} \nonumber \\
&& \hphantom{\sum_n\biggl[}+ \frac{
\langle\psi_0 | \hat{B} | \psi_n\rangle \langle \psi_n | \hat{A} |\psi_0\rangle}{
E_n-E_0+\omega-{\rm i}\eta}
\biggr] \; , \\
 \tilde{G}_{A,B}^{\rm ret}(\omega) &=&
\sum_n \biggl[ 
\frac{
\langle \psi_0 | \hat{A} | \psi_n\rangle \langle \psi_n | \hat{B}|\psi_0\rangle}{
E_0-E_n+\omega+{\rm i}\eta} \nonumber \\
&& \hphantom{\sum_n\biggl[}
+ \frac{
\langle \psi_0 | \hat{B} | \psi_n\rangle \langle \psi_n |\hat{A}|\psi_0\rangle}{
E_n-E_0+\omega+{\rm i}\eta}
\biggr] \; .
\label{eq:Lehmannret}
\end{eqnarray}
The Lehmann representation shows that the real parts of the causal and
retarded Green function agree and that their imaginary parts
differ in sign for $\omega<0$. 
Therefore, we can derive the causal Green function from the
retarded Green function by the simple substitution
\begin{equation}
 \tilde{G}_{A,B}^{\rm c}(\omega) =  
\left. \tilde{G}_{A,B}^{\rm ret}(\omega)\right|_{\omega+{\rm i}\eta \to 
\omega+{\rm i}{\rm sgn}(\omega)\eta}
\label{eq:retardedtocausal}
\end{equation}
in frequency space where ${\rm sgn}(u)=\Theta(u)-\Theta(-u)$ 
is the sign function.

\subsection{Spectral function and density of states}

Finally, we define the spectral function for the Fermion Green function
as
\begin{equation}
D_{A,B}(\omega) = -\frac{1}{\pi} {\rm Im}\left(\tilde{G}_{A,B}^{\rm ret}(\omega)
\right) \; .
\label{eq:specfuncdef}
\end{equation}
The Lehmann representation shows that it is positive semi-definite
if $\hat{A}=\hat{B}^+$.

When we use the operators for single-particle eigenstates of $\hat{H}_0$
with eigenenergies~$E(m)$,
$\hat{A}=\hat{a}_{m,\sigma}^{\vphantom{+}}$ and
$\hat{B}=\hat{a}_{m,\sigma}^+$, 
see chapter~\ref{sec:eigen},
we find from the Lehmann representation
\begin{equation}
D_{\sigma}(\omega)=\frac{1}{L}
\sum_m D_{m,\sigma;m,\sigma}(\omega)=\frac{1}{L}\sum_m \delta(\omega-E(m)) 
\label{eq:dosdefgeneral}
\end{equation}
because we have $E_m=E(m)+E_0$ ($E_m=-E(m)+E_0$) for a single-particle
(single-hole) excitation of the ground state for non-interacting particles. 
Apparently, $D_{\sigma}(\omega)$ describes the density of states
for single-particle excitations with spin~$\sigma$.

\section{Solution of the Schr\"odinger equation}
\label{sec:eigen}

We analyze the non-interacting Hamiltonian $\hat{H}_0
=\hat{T}+\hat{V}$ for large but finite system sizes~$L$.
We shall take the thermodynamic limit, $L\to \infty$, where appropriate.

\subsection{Derivation of the eigenvalue equation}

Since $\hat{H}_0$ poses a single-particle problem, we may write
\begin{equation}
\hat{H}_0= \sum_{m=0,\sigma}^{L} 
E(m) \hat{a}_{m,\sigma}^+\hat{a}_{m,\sigma}^{\vphantom{+}}
\; ,
\label{eq:eigenvaluecondition}
\end{equation}
where
\begin{equation}
\hat{a}_{m,\sigma}^+ 
= g_m^*\hat{d}_{\sigma}^+ +
\sqrt{\frac{1}{L}} \sum_{n=0}^{L-1} \lambda_m^*(n) 
\hat{c}_{n,\sigma}^+
\label{eq:ainoriginals} 
\end{equation}
is the Fermion creation operator for an exact eigenmode with energy $E(m)$.
The $L+1$ energies are labeled in ascending order, $E(m-1)< E(m)$,
$m=1,2,\ldots,L$.

The orthonormality condition 
\begin{equation}
\left[ \hat{a}_{m,\sigma}^+, \hat{a}_{m',\sigma'}^{\vphantom{+}} \right]_+ = 
\delta_{\sigma,\sigma'}\delta_{m,m'}
\end{equation} 
implies
\begin{equation}
1 = | g_m|^2+ 
\frac{1}{L} \sum_{n=0}^{L-1} |\lambda_m(n)|^2 \; .
\label{eq:normalizeit}
\end{equation}
Equation~(\ref{eq:eigenvaluecondition}) can only hold if
\begin{equation}
\left[ \hat{H}_0, \hat{a}_{m,\sigma}^+  \right]_{-} = E(m) \hat{a}_{m,\sigma}^+ \; .
\label{eq:eigen}
\end{equation}
To express this equation in terms of the original operators,
see equation~(\ref{eq:ainoriginals}),
we use
\begin{eqnarray}
\left[ \hat{T}, \hat{a}_{m,\sigma}^+  \right]_{-} &=& 
\sqrt{\frac{1}{L}} \sum_{n=0}^{L-1} \lambda_m^*(n) \epsilon(n)
\hat{c}_{n,\sigma}^+ \; , \\
\left[ \hat{V}, \hat{a}_{m,\sigma}^+  \right]_{-} &=&
 \sqrt{\frac{1}{L}} g_m^* \sum_{n=0}^{L-1} V_n \hat{c}_{n,\sigma}^+ 
+\left(\frac{1}{L} \sum_{n=0}^{L-1} V_n\lambda_m^*(n)\right) \hat{d}_{\sigma}^+
\; .\nonumber \\
\end{eqnarray}
A comparison with equation~(\ref{eq:eigen}) leads to the conditions
\begin{eqnarray}
E(m) g_m^* &=&  \frac{1}{L} \sum_{n=0}^{L-1} V_n \lambda_m^*(n)  \; ,\\
E(m) \lambda_m^*(n) &=&  \epsilon(n)\lambda_m^*(n)
+ g_m^*V_n^*
\; .
\end{eqnarray}
We thus find
\begin{equation}
\lambda_m(n)  = g_m \frac{V_n}{E(m)-\epsilon(n)} 
\end{equation}
with the energies from the eigenenergy equation~\cite{Anderson1961,Solyombook}
\begin{equation}
E(m)  = \frac{1}{L} \sum_{n=0}^{L-1} \frac{|V_n|^2}{E(m)-\epsilon(n)}\; .
\label{eq:eigenenergyequation}
\end{equation}
The solutions of the eigenenergy equation 
provide all the information about the finite-size system.
The normalization condition~(\ref{eq:normalizeit}) reduces to
\begin{equation}
| g_m|^2  = |g(E(m))|^2
= \left(
1+ \frac{1}{L} \sum_{n=0}^{L-1} \frac{|V_n|^2}{(E(m)-\epsilon(n))^2}
\right)^{-1} \; .
\label{eq:normalizeagain}
\end{equation}

\begin{figure}[ht]
\vspace*{5pt}
\includegraphics[width=\columnwidth]{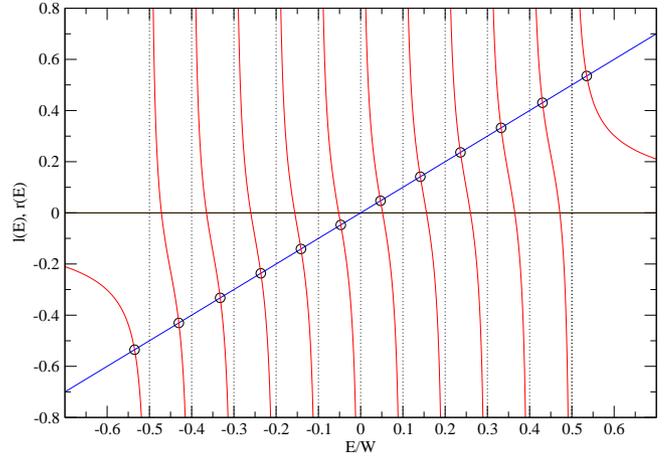}%
\vspace*{-5pt}
\caption{\label{fig:first}\col Graphical solution of the eigenenergy 
equation~(\protect\ref{eq:eigenenergyequation})
for $L=11$, $V_n^2/L=0.01$, and a linear dispersion relation,
$\epsilon(n) = -1/2+n/(L-1)$.
The left-hand-side of equation~(\protect\ref{eq:eigenenergyequation}),
$l(E)=E$,
is shown as a straight, blue line, the 
right-hand-side of equation~(\protect\ref{eq:eigenenergyequation}),
$r(E)=\sum_{n=0}^{L-1}(V_n^2/L)/[E-\epsilon(n)]$,
is shown by the red lines. Their intersections, $l(E(m))=r(E(m))$,
the eigenenergies, are encircled.
For $E<-1/2$ ($E>1/2$) we find the (anti-)bound states,
the scattering states lie in the interval $|E|<1/2$.
The vertical dotted lines indicate the 
divergences of $r(E)$ at $E=\epsilon(n)$, $n=0,\ldots,L-1$.}
\end{figure}

As an example, 
in figure~\ref{fig:first}
we show the graphical solution of the 
eigenenergy equation~(\ref{eq:eigenenergyequation}) 
for $L=11$ and $V_n^2/L=0.01$ and a linear dispersion relation,
$\epsilon(k) = -1/2+k/(L-1)$ (bandwidth $W=1$).
The figure displays particle-hole symmetry, and bound/anti-bound states
as well as scattering states, as we discuss in the remainder of this section.

\subsection{Particle-hole symmetry}

If $E(m)$ is a solution of the eigenenergy 
equation~(\ref{eq:eigenenergyequation}), 
$[-E(m)]$ also is a solution.
This is easily shown with the help of particle-hole symmetry,
\begin{eqnarray}
\frac{1}{L} \sum_{n=0}^{L-1} \frac{|V_n|^2}{-E(m)-\epsilon(n)}
&=& - \frac{1}{L} \sum_{n=0}^{L-1} \frac{|V_n|^2}{E(m)+\epsilon(n)} 
\nonumber \\
&=& 
- \frac{1}{L} \sum_{n=0}^{L-1} \frac{|V_{L-1-n}|^2}{E(m)+\epsilon(L-1-n)} 
\nonumber \\
&=& 
- \frac{1}{L} \sum_{n=0}^{L-1} \frac{|V_n|^2}{E(m)-\epsilon(n)}
= -E(m) \; ,
\end{eqnarray}
where we used the symmetry conditions
$V_{n}=V_{L-1-n}^*$ and $\epsilon(n)=-\epsilon(L-1-n)$.
Therefore, using our energy labeling,
we find $E(L-m)=-E(m)$ for $m=0,1,\ldots,L$.

\subsection{Bound/anti-bound states}

Outside the band edges, bound and anti-bound states can form.
In the thermodynamic limit, their energies $E_{\rm b}<-1/2$
($E_{\rm ab}>1/2$) are obtained from the
solution of the integral equation
\begin{equation}
E_{\rm b/ab}= \int_{-1/2}^{1/2}{\rm d} \epsilon \rho_0(\epsilon) 
\frac{|V(\epsilon)|^2}{E_{\rm b/ab}-\epsilon} \; .
\label{eq:boundantibound}
\end{equation}
Their existence depends 
on the shape of the host density of states $\rho_0(\epsilon)$
and of the hybridization $V(\epsilon)$. If the density of states
continuously goes to zero at the band edges and the hybridization
is well-behaved, there are no bound/anti-bound states
in the limit of small hybridization, $|V(\epsilon)|^2\ll 1$. 

For example, when we use the semi-elliptic density of states~(\ref{eq:rhosemiellipse})
and a constant hybridization in equation~(\ref{eq:boundantibound}) we find
the condition
\begin{equation}
\frac{1}{4V^2}=2-\sqrt{4-\frac{1}{(E_{\rm b/ab}^{\rm se})^2}} \; .
\end{equation}
For $V<1/4$, the semi-elliptic density of states 
does not support bound/anti-bound states.
For $V>1/4$, the bound/anti-bound levels lie at
$E_{\rm b/ab}^{\rm se}=\pm 4V^2/\sqrt{16V^2-1}$.
In contrast, for a constant density of states and a constant hybridization,
equation~(\ref{eq:boundantibound}) leads to
\begin{equation}
E_{\rm a/ab}^{\rm cons}=V^2\ln\left| 
\frac{1+2E_{\rm a/ab}^{\rm cons}}{1-2E_{\rm a/ab}^{\rm cons}}
\right|\; .
\end{equation}
For small $V<1/4$, the bound/anti-bound levels lie at
$E_{\rm b/ab}^{\rm cons}\approx \pm (1/2 +\exp[-1/(2V^2)])$.
The (anti-)binding energy is exponentially small but 
finite for small~$V$.

The existence of (anti-)bound states
influences the energy levels in the vicinity of the band edges.
Although these effects often are negligibly small, in the following we restrict ourselves
to situations where bound/anti-bound states are absent as for
the semi-elliptic density of states for a constant, small hybridization~$V<1/4$.

\subsection{Scattering states}
\label{sec:howtocalculatescatteringstates}

For all other states, the impurity scattering induces energy shifts 
of the order of $1/L$.
Therefore, in equation~(\ref{eq:eigenenergyequation}) we set
\begin{equation}
E(m)=\epsilon(m) +\frac{x(\epsilon(m))}{L}
\quad , \quad |x(\epsilon(m))| = {\cal O}\left( 1 \right)\; ,
\label{eq:introducex}
\end{equation}
where $x(\epsilon(m))$ quantifies the scattering
energy shift introduced by the impurity.
Note that $x(\epsilon)<0$  ($x(\epsilon)>0$) for $\epsilon<0$  ($\epsilon>0$)
because  the impurity level at energy $\epsilon=0$
repels the host energy levels. We shall show that $x(0^+)=1/(2\rho_0(0))$
so that $x(\epsilon)$ is discontinuous at $\epsilon=0$.

In order to solve the eigenvalue equation~(\ref{eq:eigenenergyequation})
for large systems we start with the observation
that the Taylor expansion for finite $r\ll L$ leads to
the following approximation 
\begin{equation}
L(\epsilon(m+r)-\epsilon(m)) \approx r f'\left(f^{-1}(\epsilon(m))\right) 
= \frac{r}{\rho_0(\epsilon(m))}
\; ,
\label{eq:Taylorme}
\end{equation}
with corrections of the order $1/L$, 
see equations~(\ref{eq:defineepsasffunction}),~(\ref{eq:defrhozero}).
In the limit of large system size and not infinitesimally close
to the band edges, we can write
{\arraycolsep=0.7pt\begin{eqnarray}
\sum_{n=0}^{L-1} \frac{|V(\epsilon(n))|^2/L}{E(m)-\epsilon(n)}
&=& \lim_{R\to\infty} \sum_{r=-R}^{R} \!
\frac{|V(\epsilon(m+r))|^2}{x(\epsilon(m))-L(\epsilon(m+r)-\epsilon(m))}
\nonumber \\
&& + \Lambda_V(E(m)) \; ,
\label{eq:summeup}
\end{eqnarray}}%
where 
\begin{equation}
\Lambda_V(E) = {\cal P}\int_{-1/2}^{1/2}{\rm d}\epsilon \rho_0(\epsilon)
\frac{|V(\epsilon)|^2 }{E-\epsilon} \; ,
\end{equation}
and ${\cal P}$ denotes the Cauchy principal value integral.
For constant hybridization and the semi-elliptic density of states we have
for $|E|<1/2$
\begin{equation}
\Lambda_V^{\rm se}(E) = 8 V^2 E\; .
\label{eq:Lambdasemiellipse}
\end{equation}
This particularly simple form permits explicit calculations, see below.

For the derivation of equation~(\ref{eq:summeup}) we 
singled out the region $|m-n|\leq R$ ($1\ll R\ll L$) 
from the sum over~$n$ 
before we employed the Euler-Maclaurin
sum formula,
\begin{equation}
\sum_{n=N_a}^{N_b} h(n) = \int_{N_a}^{N_b} {\rm d} n h(n)
+\frac{1}{2} \left( h(N_a)+h(N_b)\right) + \ldots 
\label{eq:EulerMaclaurin}
\end{equation}
that generates the contribution $\Lambda_V(E)$ in equation~(\ref{eq:summeup}). 
For the first term in equation~(\ref{eq:summeup}) 
we use equation~(\ref{eq:Taylorme}) 
{\arraycolsep=1pt\begin{eqnarray}
\sum_{r=-\infty}^{\infty} 
\frac{|V(\epsilon(m+r))|^2}{x(\epsilon(m))-L(\epsilon(m+r)-\epsilon(m))}
\approx
|V(\epsilon(m))|^2\rho_0(\epsilon(m)) && \nonumber \\
\times \sum_{r=-\infty}^{\infty} \frac{1}{x(\epsilon(m))\rho_0(\epsilon(m))-r}\; .
&&\nonumber \\
\end{eqnarray}}%
Using equation~(1.421,3) of Ref.~\cite{GR} we find
{\arraycolsep=0.4pt\begin{eqnarray}
\sum_{r=-\infty}^{\infty} 
\frac{|V(\epsilon(m+r))|^2}{x(\epsilon(m))-L(\epsilon(m+r)-\epsilon(m))}
\approx \pi \rho_0(\epsilon(m)) |V(\epsilon(m)|^2 &&\nonumber \\
\times \cot\left[\pi \rho_0(\epsilon(m))x(\epsilon(m))\right] \; . &&\nonumber \\
\label{eq:cotangens}
\end{eqnarray}}%
Here we used the fact that $V(\epsilon)$ is a smooth function so that
$|V(\epsilon(m+r))|^2\approx |V(\epsilon(m))|^2+{\cal O}(1/L)$.
To leading order in $1/L$, the eigenvalue 
equation~(\ref{eq:eigenenergyequation}) leads to
\begin{equation}
x(\epsilon) = \frac{1}{\pi\rho_0(\epsilon)}
\cot^{-1}\left[\frac{\epsilon-\Lambda_V(\epsilon)}{
\pi |V(\epsilon)|^2\rho_0(\epsilon)}\right] \; .
\label{eq:xfinalresult}
\end{equation}
This is the desired equation for the scattering energy shifts;
for a constant density of states, the derivation can be found as
equation~(I-12) in Ref.~\cite{Schoeller1998}.

For later use, we define
\begin{eqnarray}
x(\epsilon) &=& \frac{1}{2\rho_0(\epsilon)}
\left( \Theta(\epsilon)-\Theta(-\epsilon) \right)
- y(\epsilon) \; , \nonumber \\
y(\epsilon)&=&
\frac{1}{\pi \rho_0(\epsilon)}
\tan^{-1}\left[\frac{\epsilon-\Lambda_V(\epsilon)}{
\pi |V(\epsilon)|^2\rho_0(\epsilon)}\right] \; .
\label{eq:yintroduction}
\end{eqnarray}
Note that, for a smooth hybridization $V(\epsilon)$, the function $y(\epsilon)$ 
is continuous in the interval $|\epsilon|\leq 1/2$.

\section{Ground-state expectation values}
\label{sec:gsexpectationvalues}

According to equation~(\ref{eq:eigenvaluecondition})
the ground state of $\hat{H}_0$ is given by
\begin{equation}
|\psi_0\rangle = \prod_{\sigma} \prod_{m=0}^{(L-1)/2}\hat{a}_{m,\sigma}^+
| \hbox{vac}\rangle \; .
\end{equation}

\subsection{Ground-state energy}

We are interested in the change  $\Delta E$ of the ground-state energy
due to the hybridization of the impurity and the host electrons.
In the absence of bound/anti-bound states, 
it is given by
\begin{equation}
\frac{\Delta E}{2} =
\sum_{m=0}^{(L-1)/2} \left[E(m)-\epsilon(m)\right] \; .
\end{equation}
Here we took into account that the $(L+1)/2$ states lowest in energy
are occupied for each spin species. Moreover, $\epsilon((L-1)/2)=0$
and the impurity level is at $E_{\rm d}=0$ so that they do not contribute
in the case of vanishing hybridization.

The Euler-Maclaurin formula~(\ref{eq:EulerMaclaurin}) and the definition
of the host density of states~(\ref{eq:defrhozero}) lead to
\begin{equation}
\frac{\Delta E}{2} =
\int_{-1/2}^0{\rm d}\epsilon \rho_0(\epsilon) x(\epsilon)
= \int_{-1/2}^0\frac{{\rm d}\epsilon }{\pi} 
\cot^{-1}\left[\frac{\epsilon-\Lambda_V(\epsilon)}{
\pi |V(\epsilon)|^2\rho_0(\epsilon)}\right] 
\label{eq:energyintermediate}
\end{equation}
in the thermodynamic limit where we inserted the scattering energy shifts from 
equation~(\ref{eq:xfinalresult}).

Equation~(\ref{eq:energyintermediate}) can be evaluated further in the limit of
vanishingly small hybridization. We set $V(\epsilon)=V v(\epsilon)$ with
$v(0)=1$ and consider $V \to 0$. Then,
\begin{equation}
\frac{\Delta E}{2} \approx
\frac{V^2}{\pi} \int_{-1/2}^{-c V^2}{\rm d}\epsilon 
\frac{\pi |v(\epsilon)|^2\rho_0(\epsilon)}{\epsilon}
\label{eq:energysmallV}
\end{equation}
with a low-energy cut-off, $c={\cal O}(1)$.
To leading order in $V^2\ln(1/V^2)$ we then find
\begin{equation}
\frac{\Delta E}{2} (V\to 0) =
- \rho_0(0) V^2\ln\left(\frac{1}{V^2}\right) +{\cal O}\left(V^2\right) \; .
\label{eq:energysmallVfinal}
\end{equation}
For a constant hybridization and the semi-elliptic density of states we find
for all $|V|<1/4$
\begin{equation}
\frac{(\Delta E)^{\rm se}(V)}{2} =
-\frac{\alpha}{2\pi}
\frac{\tanh^{-1}\left[\sqrt{1-\alpha^2}\right]}{\sqrt{1-\alpha^2}}
\quad , \quad \alpha= \frac{8V^2}{1-8V^2} \;.
\label{eq:energysmallVfinalsemi}
\end{equation}
For small $V$ this can be approximated as ($\ln(e)=1$)
\begin{eqnarray}
\frac{(\Delta E)^{\rm se}(V)}{2} &=&
-\frac{4}{\pi}V^2 \left[ \ln\left(\frac{1}{4V^2}\right)
+8V^2\ln\left(\frac{1}{4 e V^2}\right)  
\right] \nonumber \\
&& +{\cal O}\left(V^6\ln(V^2) \right) \nonumber\\
&=&-\frac{\Gamma}{\pi}\left[ \ln\left(\frac{1}{\Gamma}\right)
+8V^2\ln\left(\frac{1}{e\Gamma}\right)  
\right] +{\cal O}\left(V^6\ln(V^2) \right) 
\nonumber \; , \\
\Gamma &=& \pi \rho_0(0) V^2 
\label{eq:energysmallVfinalsemiGamma}
\; .
\end{eqnarray}
For a constant hybridization and a constant density of states,
the small-$V$ expansion of the ground-state energy shift 
is given by
\begin{eqnarray}
\frac{(\Delta E)^{\rm cons}(V)}{2} &=&
-V^2 \left[ \ln\left(\frac{e}{2\pi V^2}\right)
+4 V^2\ln\left(\frac{1}{V^2}\right)  \right]
+{\cal O}\left(V^4 \right) \nonumber\\
&=&-\frac{\Gamma}{\pi}\left[ \ln\left(\frac{e}{2 \Gamma}\right)
+4V^2\ln\left(\frac{\pi}{\Gamma}\right)  
\right] +{\cal O}\left(V^4\right) 
\nonumber \label{eq:energysmallVfinalconsGamma}
\; .
\end{eqnarray}
The comparison with the general low-$V$ expansion~(\ref{eq:energysmallVfinal})
shows that the correction of the order ${\cal O}(V^2)$ depends
on the shape of the host density of states. 

\subsection{Impurity occupancy}

With the help of
\begin{equation}
\left[ \hat{d}_{\sigma}^+\hat{d}_{\sigma}^{\vphantom{+}}, \hat{a}_{m,\sigma'}^+
\right]_{-} = \delta_{\sigma,\sigma'} g_m^* \hat{d}_{\sigma}^+
\end{equation}
we find that
\begin{equation}
\langle \hat{d}_{\sigma}^+\hat{d}_{\sigma}^{\vphantom{+}}\rangle
= \sum_{m=0}^{(L-1)/2} |g_m|^2  \; .
\label{eq:ddirect}
\end{equation}
Equation~(\ref{eq:normalizeagain}) shows that 
$|g_{L-m}|^2=|g_m|^2$ so that
\begin{equation}
\sum_{m=0}^{L} |g_m|^2  
= 2 \sum_{m=0}^{(L-1)/2} |g_m|^2  = 
2 \langle \hat{d}_{\sigma}^+\hat{d}_{\sigma}^{\vphantom{+}}\rangle \;.
\end{equation}
The expression on the left-hand side corresponds to the
probability to find the $d$-level occupied in a completely filled
system, 
\begin{equation}
\sum_{m=0}^{L} |g_m|^2  = 1 \; .
\end{equation}
Therefore we find the result
\begin{equation}
\langle \hat{d}_{\sigma}^+\hat{d}_{\sigma}^{\vphantom{+}}\rangle = \frac{1}{2}
\label{eq:eq:dishalfphagain}
\end{equation}
as a consequence 
of particle-hole symmetry, in agreement with equation~(\ref{eq:dishalf}).

It is instructive to derive equation~(\ref{eq:eq:dishalfphagain}) explicitly. From
equation~(\ref{eq:introducex}) 
we find up to terms of ${\cal O}(1) $
\begin{equation}
\sum_{n=0}^{L-1} \frac{|V(n)|^2/L}{(E(m)-\epsilon(n))^2}
= -\frac{\partial}{\partial x(\epsilon(m))}
\left[
\sum_{n=0}^{L-1} \frac{|V(n)|^2}{E(m)-\epsilon(n)}\right] \; ,
\end{equation}
so that from equation~(\ref{eq:cotangens}) we find
\begin{equation}
\sum_{n=0}^{L-1} \frac{|V(n)|^2/L}{(E(m)-\epsilon(n))^2}
=L \frac{[\pi \rho_0(\epsilon(m)|V(\epsilon(m))|]^2
}{
\sin^2[\pi \rho_0(\epsilon(m))x(\epsilon(m))]} \; .
\end{equation}
Therefore, equations~(\ref{eq:normalizeagain}) and~(\ref{eq:xfinalresult}) give
\begin{eqnarray}
|g_m|^2 &=& \frac{1}{L} |g(\epsilon(m))|^2\nonumber \; ,\\
|g(\epsilon)|^2&=& 
\frac{|V(\epsilon)|^2}{ [\pi\rho_0(\epsilon)|V(\epsilon)|^2]^2 
+[\epsilon-\Lambda_V(\epsilon)]^2}
 \; ,
\label{eq:gepsTDL}
\end{eqnarray}
where we used equation~(\ref{eq:xfinalresult}) for $x(E)$.
Then, from equation~(\ref{eq:ddirect}) 
\begin{eqnarray}
\langle \hat{d}_{\sigma}^+\hat{d}_{\sigma}^{\vphantom{+}}\rangle
&=& \int_{-1/2}^0{\rm d}E \rho_0(E) |g(E)|^2 \nonumber \\
&=& 
\int_{-1/2}^0{\rm d}E 
\frac{\rho_0(E) |V(E)|^2}{ [\pi\rho_0(E)|V(E)|^2]^2 +[E-\Lambda_V(E)]^2}
 \; .
\label{eq:doccanalytic}
\end{eqnarray}
Using $\rho_0(-E)=\rho_0(E)$, $|V(-E)|^2=|V(E)|^2$, and
$\Lambda_V(-E)=-\Lambda_V(E)$ due to particle-hole symmetry, we can write
\begin{equation}
2 \langle \hat{d}_{\sigma}^+\hat{d}_{\sigma}^{\vphantom{+}}\rangle
= \int_{-1/2}^{1/2}{\rm d}E 
\frac{\rho_0(E) |V(E)|^2}{ [\pi\rho_0(E)|V(E)|^2]^2 +[E-\Lambda_V(E)]^2}
= 1
\label{eq:doccanalytic2}
\end{equation}
because the integral in equation~(\ref{eq:doccanalytic2})
gives the result for a completely filled band. Therefore, we find
$\langle \hat{d}_{\sigma}^+\hat{d}_{\sigma}^{\vphantom{+}}\rangle=1/2$
again.

\subsection{Hybridization}

With the help of
\begin{equation}
\left[ \hat{c}_{k,\sigma}^+\hat{d}_{\sigma}^{\vphantom{+}}, \hat{a}_{m,\sigma'}^+
\right]_{-} = \delta_{\sigma,\sigma'} g_m^* \hat{c}_{k,\sigma}^+
\end{equation}
we find that
\begin{eqnarray}
\langle \hat{c}_{k,\sigma}^+\hat{d}_{\sigma}^{\vphantom{+}}\rangle
&=& \sqrt{\frac{1}{L}}
\sum_{m=0}^{(L-1)/2} g_m^* \lambda_m(k)  \nonumber \\
&=& \sqrt{\frac{1}{L}}\sum_{m=0}^{(L-1)/2} |g_m|^2 \frac{V_k}{E(m)-\epsilon(k)} 
\; .
\end{eqnarray}
In the thermodynamic limit, this expression can be transformed into
\begin{eqnarray}
\langle \hat{c}_{k,\sigma}^+\hat{d}_{\sigma}^{\vphantom{+}}\rangle
&\equiv& \frac{V_k}{\sqrt{L}}\left[ G(\epsilon(k)) + H(\epsilon(k))\right]  
\; , \label{eq:hybelement}\\
G(\epsilon) &=& \Theta(-\epsilon)| g(\epsilon)|^2 
\frac{\left[\epsilon -\Lambda_V(\epsilon)\right]}{|V(\epsilon)|^2}
\nonumber\\
&=& \Theta(-\epsilon)
\frac{\epsilon -\Lambda_V(\epsilon)}{
\left[\pi\rho_0(\epsilon)|V(\epsilon)|^2\right]^2
+\left[\epsilon -\Lambda_V(\epsilon)\right]^2} \; ,
\label{eq:Gintegral}\\
H(\epsilon) &=& \int_{-1/2}^0 \frac{{\rm d}E}{E-\epsilon} 
\frac{\rho_0(E) |V(E)|^2}{ [\pi\rho_0(E)|V(E)|^2]^2 +[E-\Lambda_V(E)]^2}
\; ,\nonumber \\
\label{eq:Hintegral}
\end{eqnarray}
where the integral on the right-hand side of equation~(\ref{eq:Hintegral})
must be understood as
a principal value integral when $-1/2<\epsilon<0$.
The derivation of $G(\epsilon)$ proceeds along the lines developed
in Sect.~\ref{sec:howtocalculatescatteringstates}.

In general, $H(\epsilon)$ cannot be calculated analytically.
In the limit of vanishing hybridization,
$V(\epsilon)=V v(\epsilon)$ with $v(0)=1$ and $V\to 0$, we find
\begin{eqnarray}
H(\epsilon,V\to 0) &\approx&
\int_{-\infty}^0 {\rm d}E 
\frac{V^2\rho_0(0)}{ [\pi \rho_0(0)V^2]^2 +E^2}\frac{1}{E-\epsilon} \nonumber \\
&=& -\frac{\epsilon}{ 2[(\pi \rho_0(0)V^2)^2 +\epsilon^2]}
\nonumber \\
&& + \rho_0(0)V^2 
\frac{\ln\left(|\epsilon|/(\pi \rho_0(0)V^2)\right)}{ 
(\pi \rho_0(0)V^2)^2 +\epsilon^2}
\; .
\label{eq:HforsmallV}
\end{eqnarray}
Apparently, the hybridization matrix element is logarithmically divergent
near $\epsilon=0$. 
This does not cause any problems because 
$\langle \hat{c}_{k,\sigma}^+\hat{d}_{\sigma}^{\vphantom{+}}\rangle
\sim [G(\epsilon(k))+H(\epsilon(k))]/\sqrt{L}$ remains bounded 
since the smallest accessible values for $\epsilon(k)$ is of
the order of $1/L$.
For a constant hybridization and the semi-elliptic density
of states, $H^{\rm se}(\epsilon)$ can be calculated
analytically. The lengthy expressions agree very well with 
$H(\epsilon,V\to 0)$ for all $V<0.1$.

The contribution of the hybridization to the ground-state energy
per spin is given by
\begin{equation}
\frac{\langle\hat{V}\rangle}{2} 
= 2 \int_{-1/2}^{1/2} {\rm d}\epsilon \rho_0(\epsilon)
|V(\epsilon)|^2 \left[H(\epsilon)+G(\epsilon)\right] \; .
\end{equation}
In the limit of vanishingly small hybridization,
$H(\epsilon)$ does not contribute to the hybridization 
energy. The first term of $H(\epsilon)$ in equation~(\ref{eq:HforsmallV})
is odd and thus cancels out when integrated over the whole band.
The second term apparently is of the order $V^4\ln(V^2)$
and thus smaller by a factor of $V^2$ than the leading-order term.
Therefore, we have
\begin{eqnarray}
\frac{\langle\hat{V}\rangle}{2}(V\to 0) 
&\approx& 2 \int_{-cV^2}^0 {\rm d}\epsilon 
\frac{\rho_0(0)V^2\epsilon}{\left[\pi \rho_0(0)V^2\right]^2+\epsilon^2}
\nonumber \\
&=& -2 \frac{\Gamma}{\pi}\ln\left(\frac{1}{\Gamma}\right) 
+ {\cal O}\left(V^2\right) = 2 \frac{\Delta E}{2}(V\to 0)\; ,
\label{eq:hybridizationenegry}
\end{eqnarray}
see equation~(\ref{eq:energysmallVfinalsemiGamma}).
The energy gain through the hybridization is twice as large as
the energy loss due to the distortion of the Fermi sea, as we show next.

\subsection{Momentum distribution}

With the help of
\begin{equation}
\left[ \hat{c}_{k,\sigma}^+\hat{c}_{k,\sigma}^{\vphantom{+}}, 
\hat{a}_{m,\sigma'}^+
\right]_{-} = \delta_{\sigma,\sigma'} \sqrt{\frac{1}{L}} 
\lambda_m^*(k) \hat{c}_{k,\sigma}^+
\end{equation}
we find that
\begin{eqnarray}
\langle \hat{c}_{k,\sigma}^+\hat{c}_{k,\sigma}^{\vphantom{+}}\rangle
&=& \frac{1}{L}\sum_{m=0}^{(L-1)/2} |\lambda_m(k) |^2\nonumber \\
&=& \frac{1}{L}\sum_{m=0}^{(L-1)/2} |g_m|^2 \frac{|V_k|^2}{(E(m)-\epsilon(k))^2} 
\; .
\end{eqnarray}
The thermodynamic limit is more subtle than for the
hybridization matrix element because terms of order
unity appear next to terms of order $1/L$.
Proceeding along the lines of Sect.~\ref{sec:howtocalculatescatteringstates}
we find to leading order
\begin{equation}
\langle \hat{c}_{k,\sigma}^+\hat{c}_{k,\sigma}^{\vphantom{+}}\rangle^{(0)}
\equiv n^{(0)}(\epsilon(k))
= \Theta\left(-\epsilon(k)\right) \; .
\end{equation}
The $1/L$ corrections are obtained as
\begin{eqnarray}
L\langle \hat{c}_{k,\sigma}^+\hat{c}_{k,\sigma}^{\vphantom{+}}\rangle^{(1)}
&\equiv& n^{(1)}(\epsilon(k))\nonumber \\
&=&
|V_k|^2 \frac{\partial}{\partial u} 
\left[\sum_{m=0}^{(L-1)/2} |g_m|^2 \frac{1}{E(m)-u} \right]_{u=\epsilon(k)}
\nonumber \\
&=& |V_k|^2 \left[G'(\epsilon(k))+H'(\epsilon(k))\right]
\label{eq:momdistrfinite}
\end{eqnarray}
with $G(\epsilon)$ and $H(\epsilon)$ 
from equations~(\ref{eq:Gintegral}) and~(\ref{eq:Hintegral}), respectively.
The total momentum distribution is given by
$n(\epsilon(k))=n^{(0)}(\epsilon(k))+n^{(1)}(\epsilon(k))/L$. 

Our analysis of the function $H(\epsilon)$ in
the previous subsection shows that the momentum distribution 
develops a $1/\epsilon$ singularity for $\epsilon\to 0$
because $H(\epsilon)\sim \ln(|\epsilon|)$ 
so that $H'(\epsilon)\sim 1/\epsilon$ for $\epsilon\to 0$.
However, its strength is proportional to $V^4/L$ for small~$V$ so
that, for the smallest accessible value for $\epsilon(k)$, 
the contribution to the momentum distribution
actually remains small.

The contribution of the host electrons to the ground-state energy is given by
\begin{eqnarray}
\frac{\Delta T}{2} = \frac{\langle \hat{T}\rangle - T_0}{2}&=&
\int_{-1/2}^{1/2} {\rm d}\epsilon \epsilon \rho_0(\epsilon)
L\left(n(\epsilon)-\Theta(-\epsilon)\right)\nonumber \\
&=& \int_{-1/2}^{1/2} {\rm d}\epsilon 
\epsilon \rho_0(\epsilon)|V(\epsilon)|^2\left[G'(\epsilon)+H'(\epsilon)\right]\; ,
\nonumber \\
\end{eqnarray}
where 
\begin{equation}
T_0=2L\int_{-1/2}^{0} {\rm d}\epsilon \epsilon \rho_0(\epsilon)
\end{equation}
is the energy of the undisturbed host band.
As for the hybridization energy, the function $G(\epsilon)$
gives the dominant contribution in the limit of small hybridizations.
We find after a partial integration
\begin{eqnarray}
\frac{\Delta T}{2}(V\to 0) &\approx& 
V^2\rho_0(0)\int_{-cV^2}^{0} {\rm d}\epsilon G(\epsilon)\nonumber \\
&=& V^2 \rho_0(0)\int_{-cV^2}^{0} {\rm d}\epsilon 
\frac{\epsilon}{(\pi \rho_0(0)V^2)^2+\epsilon^2} \nonumber \\
&=&
\frac{\Gamma}{\pi}\ln\left(\frac{1}{\Gamma}\right) =-\frac{E_0}{2}(V\to 0)
\; ,
\label{eq:hostenergy}
\end{eqnarray}
including only the leading-order terms, of the order of
${\cal O}\left(V^2\ln(1/V^2)\right)$.
Equation~(\ref{eq:hostenergy}) shows that, indeed, the host electrons' loss
in energy is half of the gain due to their hybridization with
the impurity, compare equation~(\ref{eq:hybridizationenegry}).

\section{Spectral properties}
\label{sec:eigenspec}

In this chapter we derive the single-particle spectral pro\-perties.
In the supplementary material, we use the equa\-tion-of-motion approach
to derive the Green functions and ground-state expectation values.

\subsection{Density of states}
\label{sec:densityofstates}

We start with the density of states for the system without hybridization.
It is given by
\begin{eqnarray}
D_0(\omega) &=& D_{{\rm imp},0}(\omega)+D_{\rm host}(\omega)\nonumber\;, \\ 
D_{{\rm imp},0}(\omega)&=&  \delta(\omega) \nonumber \; ,\\
D_{\rm host}(\omega)&=& \sum_{n=0}^{L-1} \delta\left(\omega-\epsilon(n)\right)
\; ,
\end{eqnarray}
where we simply 
added the contributions from the impurity and the host electrons, 
see equation~(\ref{eq:dosdefgeneral}).
Altogether there are $L+1$ energy levels,
\begin{equation}
\int_{-\infty}^{\infty}{\rm d}\omega D_0(\omega) = L+1\; .
\end{equation}
Using the Euler-Maclaurin formula~(\ref{eq:EulerMaclaurin})
we readily find in the thermodynamic limit
\begin{eqnarray}
D_{\rm host}(\omega) &=& 
\int_{0}^{L-1}
{\rm d}n \delta\left(\omega-\epsilon(n)\right) \nonumber \\
&& + \frac{1}{2} 
\left[ \delta(\omega-\epsilon(0)) +\delta(\omega-\epsilon(L-1)\right]
\nonumber \\
&=& (L-1)\rho_0(\omega) 
+ \frac{1}{2} \left[ \delta(\omega+1/2) +\delta(\omega-1/2)\right]
\nonumber \; .
\end{eqnarray}
Note that in $D_0(\omega)$ we have to keep all corrections to order unity.

{}From equation~(\ref{eq:dosdefgeneral}) we have for finite hybridization
\begin{equation}
D_{\sigma}(\omega) =  \sum_{m=0}^{L} \delta\left(\omega-E(m)\right)
\; .
\end{equation}
The same steps as above lead to
\begin{eqnarray}
D_{\sigma}(\omega) &=&  
\delta(\omega) 
+ \frac{1}{2} \left[ \delta(\omega+1/2) +\delta(\omega-1/2)\right]\nonumber \\
&& + (L-1) \left( \int_{-1/2}^{0^-} + \int_{0^+}^{1/2} \right) 
{\rm d}\epsilon \rho_0(\epsilon) \delta(E(\epsilon)-\omega) 
\; ,
\end{eqnarray}
where we took special care of the step discontinuity of $x(\epsilon)$
at $\epsilon=0$, see equation~(\ref{eq:yintroduction}).
Now that
\begin{equation}
E(\epsilon) =  \epsilon+ \frac{x(\epsilon)}{L} \quad, \quad
{\rm d}\epsilon= {\rm d}E
\left( 1-\frac{x'(\epsilon)}{L}\right)+{\cal O}\left(1/L\right)
\; ,
\end{equation}
we find up to corrections in $1/L$
\begin{eqnarray}
D_{\sigma}(\omega) &=&  (L-1) \left( \int_{-1/2}^{0^-} + \int_{0^+}^{1/2} \right) 
{\rm d}E \delta(E-\omega) 
\nonumber \\
&& \hphantom{\left( \int_{-1/2}^{0^-} + \int_{0^+}^{1/2} \right)}
\times  \rho_0(E-x(E)/L)(1-x'(E)/L)
\nonumber \\
&& +\delta(\omega) + \frac{1}{2} 
\left[ \delta(\omega+1/2 ) +\delta(\omega-1/2 )\right]
\nonumber \\
&=& (L-1)\rho_0(\omega) 
+ \frac{1}{2} 
\left[ \delta(\omega+1/2 ) +\delta(\omega-1/2 )\right]\nonumber \\
&& +\delta(\omega)
-\frac{{\rm d}}{{\rm d} \omega} \left(\rho_0(\omega)x(\omega)\right)\nonumber \\
&=& D_{\rm host}(\omega)+D_{{\rm imp},\sigma}(\omega)
\; .
\end{eqnarray}
Since $x(\omega)$ is discontinuous at $\omega=0$, we find 
from~(\ref{eq:yintroduction})
\begin{equation}
D_{{\rm imp},\sigma}(\omega)=\delta(\omega)
-\frac{{\rm d}}{{\rm d} \omega} \left(\rho_0(\omega)x(\omega)\right) 
= \frac{{\rm d}}{{\rm d} \omega} \left(\rho_0(\omega)y(\omega)\right) 
\; .
\label{eq:Dimpfinal}
\end{equation}
Apparently, the $\delta$-Peak of the uncoupled impurity level broadens into
a line of finite width.

\begin{figure}[ht]
 \includegraphics[width=\columnwidth]{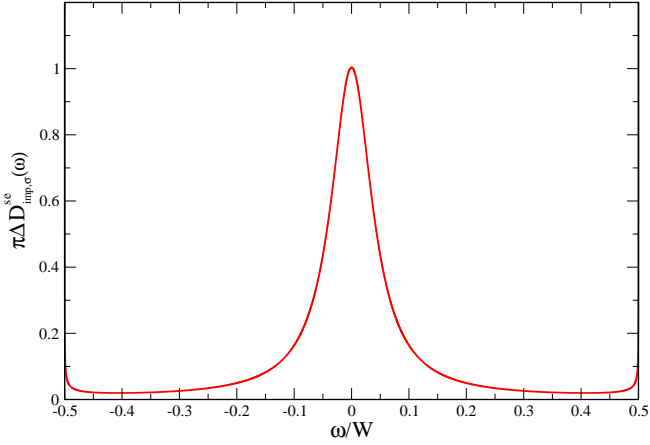}
\caption{\label{fig:second}\col Impurity spectral function for
a semi-elliptic density of states at $V=0.1W$ ($W\equiv 1$).
The dominant Lorentzian line-shape around $\omega=0$
turns into square-root divergences at the band edges.}
\end{figure}

Indeed, in the limit of small 
hybridizations~$V(\omega)=Vv(\omega)$ with $V\to 0$
and $v(0)=1$, we find from equation~(\ref{eq:Dimpfinal}) using
equation~(\ref{eq:yintroduction})
\begin{equation}
D_{{\rm imp},\sigma}(\omega)\approx
\frac{1}{\pi} \frac{\Gamma}{\omega^2 +\Gamma^2}
\quad , \quad \Gamma=\pi V^2\rho_0(0)
\; .
\end{equation}
The impurity contribution to the density of states is a Lorentzian line
of half width~$\Gamma$ at half maximum,
see equation~(\ref{eq:energysmallVfinalsemiGamma}).
For the semi-elliptic density of states and constant hybridization,
we can give an explicit result for all hybridization strengths,
\begin{eqnarray}
D_{{\rm imp},\sigma}^{\rm se}(\omega)&=&
\frac{\rho_0^{\rm se}(0)}{\rho_0^{\rm se}(\omega)}
\frac{1}{\sqrt{1-\alpha^2}}
\left(\frac{1}{\pi} \frac{\Delta}{\omega^2 +\Delta^2}\right)
\quad , \quad |\omega|<1/2\nonumber \; ,\\
&& \Delta=\frac{\alpha}{2\sqrt{1-\alpha^2}}
\quad ,\quad \alpha=\frac{8V^2}{1-8V^2} \; .
\end{eqnarray}
This example shows that the Lorentzian line shape is cut off
by the band edges. In order to guarantee the sum rule in the presence
of a finite band-width, weight accumulates close to the band edges.
In the case of the semi-elliptic density of states, 
the impurity density of states
displays square-root divergences at the band edges, see figure~\ref{fig:second}.

\subsection{Phase shift function and Friedel sum rule}

In scattering theory, the phase shift function $\eta(\epsilon)$ 
and the excess density of states are related by~\cite{Hewsonbook}
\begin{equation}
\Delta \rho(\epsilon) = \frac{1}{\pi} 
\frac{\partial \eta(\epsilon)}{\partial \epsilon}
\end{equation}
with the boundary condition $\eta(-\infty)=0$. 
In our case, $\Delta\rho(\epsilon)=D_{{\rm imp},\sigma}(\epsilon)$ and 
we see from equations~(\ref{eq:yintroduction}), (\ref{eq:Dimpfinal}) that
\begin{equation}
\eta(\epsilon) = \frac{\pi}{2}
+ \tan^{-1}\left[\frac{\epsilon-\Lambda_V(\epsilon)}{
\pi |V(\epsilon)|^2\rho_0(\epsilon)}\right] \; .
\end{equation}
This equation shows that the Friedel sum-rule is fulfilled,
$\eta(E_{\rm F})=(\pi/2)n_{\rm d}$, where $E_{\rm F}=0$ is the Fermi energy
and $n_{\rm d}=2 n_{{\rm d},\sigma}=1$ 
is the impurity occupancy at half band-filling.

\subsection{Impurity spectral function}

For non-interacting electrons, the $d$-electron Green function
is readily calculated from the Lehmann representation~(\ref{eq:Lehmannret}).
For $\hat{A}=\hat{d}_{\sigma}^{\vphantom{+}}$ 
and $\hat{B}=\hat{d}_{\sigma}^+$ only the eigenstates
$|\psi_m^{\rm p}\rangle= \hat{a}_{m,\sigma}^+|\psi_0\rangle$
and 
$|\psi_m^{\rm h}\rangle= \hat{a}_{m,\sigma}^{\vphantom{+}}|\psi_0\rangle$
contribute. We use for a particle excitation $E_m^{\rm p}=E(m)+E_0$
\begin{equation}
\langle \psi_0| \hat{d}_{\sigma}^{\vphantom{+}}
\hat{a}_{m,\sigma}^+|\psi_0\rangle = 
\langle \psi_0| \left[\hat{d}_{\sigma}^{\vphantom{+}},\hat{a}_{m,\sigma}^+\right]_+
|\psi_0\rangle = g_m^*
\; ,
\end{equation}
and likewise for a hole excitation with $E_m^{\rm h}=-E(m)+E_0$ to find
\begin{equation}
\tilde{G}_{d,d}^{\rm ret}(\omega) = \frac{1}{L}\sum_m
\frac{|g(E(m))|^2 }{\omega-E(m)+{\rm i}\eta}
\end{equation}
for the retarded $d$-electron Green function.
It is the sum over poles in the lower complex plane at the exact
excitation energies $E(m)$ with weight $|g_m|^2$.

The corresponding impurity spectral function follows from the 
definition~(\ref{eq:specfuncdef}) as
\begin{equation}
D_{d,d}(\omega) = \frac{1}{L}\sum_m |g(E(m))|^2\delta(\omega-E(m))
\; . 
\label{eq:impurityDOS}
\end{equation}
To get further insight into the spectral function,
we reconsider the eigenenergy equation~(\ref{eq:eigenenergyequation}),
\begin{eqnarray}
E(m) &=& \widetilde{\Lambda}_V(E(m)) \; , \nonumber \\
\widetilde{\Lambda}_V(\omega) &=& \frac{1}{L} \sum_p 
\frac{|V_p|^2}{\omega-\epsilon(p)}\; , \nonumber \\
\widetilde{\Lambda}_V'(\omega)&=& 
-\left( \frac{1}{|g(\omega)|^2}-1\right)
\; ,
\label{eq:LambdaVtilde}
\end{eqnarray}
where we used equation~(\ref{eq:normalizeagain}).
In the vicinity of an eigen\-energy $E(m)=\widetilde{\Lambda}_V(E(m))$
we Taylor expand
\begin{eqnarray}
\omega-\widetilde{\Lambda}_V(\omega)
&\approx &
E(m)+(\omega-E(m))\nonumber\\
&& -\widetilde{\Lambda}_V(E(m)) 
- \widetilde{\Lambda}_V'(E(m))(\omega-E(m))\; ,
\end{eqnarray}
so that
\begin{equation}
\frac{1}{\omega-\widetilde{\Lambda}_V(\omega)}
=\frac{1}{1-\widetilde{\Lambda}_V'(E(m))}
\frac{1}{\omega-E(m)} = \frac{|g_m|^2}{\omega-E(m)}\; ,
\end{equation}
where we used equations~(\ref{eq:impurityDOS}) and~(\ref{eq:LambdaVtilde}).
Therefore, we can equally write
\begin{equation}
D_{d,d}(\omega) = -\frac{1}{\pi}
{\rm Im} \left(
\frac{1}{\omega-\widetilde{\Lambda}_V(\omega)}
\right)
\end{equation}
for the impurity density of states.
With the help of equation~(\ref{eq:gepsTDL}) we can explicitly evaluate
equation~(\ref{eq:impurityDOS}) in the thermodynamic limit,
\begin{eqnarray}
D_{d,d}(\omega) &=& \int_{-1/2}^{1/2} {\rm d}\epsilon
\rho_0(\epsilon) |g(\epsilon)|^2 \delta(\epsilon-\omega)\nonumber \\
&=& \frac{\rho_0(\omega)|V(\omega)|^2}{(\pi\rho_0(\omega)|V(\omega)|^2)^2
+(\omega-\Lambda_V(\omega))^2} \; ,
\label{eq:finalimspecfunc}
\end{eqnarray}
which is the well-known result for the impurity spectral function
for the non-interacting single-impurity Anderson model.

\section{Conclusions}
\label{sec:conclusions}

In this work we started from the eigenvalue equations
to derive ground-state properties 
for the non-interacting symmetric single-impurity Anderson model.
We derived the ground-state energy, 
the hybridization and momentum distribution functions,
and various spectral functions such as the density of states, the phase-shift function
and the impurity spectral function.
For comparison, in the supplementary material
we used the standard equation-of-motion approach
to derive the Green functions and ground-state expectation values.

For a finite host bandwidth~$W$, we demonstrate that the impurity spectral function
can display a finite weight at the band edges. For a semi-elliptic density of states 
and a constant hybridization,
we give an explicit expression for the impurity spectral function
for all hybridization
strengths $V<W/4$, where no bound and anti-bound states exists.
The usual Lorentzian spectrum is recovered in the weak-hybridization 
limit, $V/W\to 0$.

Our work closes a gap in the analytical treatment of 
the single-impurity Anderson model.
Moreover, our explicit expressions for ground-state expectation values 
will be useful for variational approaches such as the Gutzwiller wave function.

\section{Supporting information}

In the supporting information, we derive the Green functions 
for the non-interacting single-impurity Green function
from the equation-of-motion method. Using the Green functions,
we calculate the total density of states 
and ground-state expectation values.
The results agree with those obtained from the direct calculations in the 
previous sections.

\subsection{Equation-of-motion approach}

\subsubsection{Time domain}

We study the four retarded Green functions
\begin{eqnarray}
G_{k,p}^{\rm ret}(t) &=& (-{\rm i}) \Theta(t) \langle 
\left[ \hat{c}_{k,\sigma}^{\vphantom{+}}(t),\hat{c}_{p,\sigma}^+\right]_+
\rangle \; , \label{eq:defGFretarded1}\\[2pt]
G_{d,p}^{\rm ret}(t) &=& (-{\rm i}) \Theta(t) \langle 
\left[\hat{d}_{\sigma}^{\vphantom{+}}(t),\hat{c}_{p,\sigma}^+\right]_+
\rangle \; , \label{eq:defGFretarded2}\\[2pt]
G_{k,d}^{\rm ret}(t) &=& (-{\rm i}) \Theta(t) \langle 
\left[ \hat{c}_{k,\sigma}^{\vphantom{+}}(t),\hat{d}_{\sigma}^+\right]_+
\rangle \; , \label{eq:defGFretarded3}\\[2pt]
G_{d,d}^{\rm ret}(t) &=& (-{\rm i}) \Theta(t) \langle 
\left[ \hat{d}_{\sigma}^{\vphantom{+}}(t),\hat{d}_{\sigma}^+\right]_+
\rangle \; , 
\label{eq:defGFretarded4}
\end{eqnarray}
Taking the time derivative leads to
\begin{eqnarray}
{\rm i} \dot{G}^{\rm ret}_{k,p}(t) &=& \delta(t) \delta_{k,p}
+(-{\rm i}) \Theta(t) 
\langle \left[ \left[ \hat{c}_{k,\sigma}^{\vphantom{+}}(t),\hat{H}_0\right]_-,
\hat{c}_{p,\sigma}^+\right]_+\rangle \nonumber \\
&=&
\delta(t) \delta_{k,p}+ \epsilon(k) G_{k,p}^{\rm ret}(t) 
+\frac{V_k^*}{\sqrt{L}} G_{d,p}^{\rm ret}(t)  \; ,
\label{eq:GFdot1}\\
{\rm i} \dot{G}_{d,p}^{\rm ret}(t) &=& 
(-{\rm i}) \Theta(t) 
 \langle  
\left[ \left[\hat{d}_{\sigma}^{\vphantom{+}}(t),\hat{H}_0\right]_-,
\hat{c}_{p,\sigma}^+\right]_+\rangle  \nonumber \\
&=& 
\sum_k\frac{V_k}{\sqrt{L}} G_{k,p}^{\rm ret}(t) \; ,
\label{eq:GFdot2}
\end{eqnarray}

\begin{eqnarray}
{\rm i} \dot{G}_{k,d}^{\rm ret}(t) &=& 
(-{\rm i})\Theta(t)  \langle  
\left[\left[ \hat{c}_{k,\sigma}^{\vphantom{+}}(t),\hat{H}_0\right]_-,
\hat{d}_{\sigma}^+\right]_+\rangle \nonumber \\
&=& 
\epsilon(k) G_{k,d}^{\rm ret}(t) +\frac{V_k^*}{\sqrt{L}} G_{d,d}^{\rm ret}(t)  \; ,
\label{eq:GFdot3}\\
{\rm i} \dot{G}_{d,d}^{\rm ret}(t) &=& 
\delta(t) 
+ (-{\rm i}) \Theta(t) \langle 
\left[\left[ \hat{d}_{\sigma}^{\vphantom{+}}(t),\hat{H}_0\right]_-,
\hat{d}_{\sigma}^+ \right]_+\rangle  \nonumber \\
&=& \delta(t) + \sum_k\frac{V_k}{\sqrt{L}} G_{k,d}^{\rm ret}(t) \; .
\label{eq:GFdot4}
\end{eqnarray}
Here, we used the anticommutation relations of the Fer\-mi operators 
and the commutation relations
\begin{eqnarray}
\left[\hat{c}_{k,\sigma}^{\vphantom{+}},\hat{T}\right]_- 
= \epsilon(k) \hat{c}_{k,\sigma}^{\vphantom{+}} &\; , \;  & 
\left[\hat{d}_{\sigma}^{\vphantom{+}},\hat{T}\right]_- 
= 0 \nonumber \; , \\
\left[\hat{c}_{k,\sigma}^{\vphantom{+}},\hat{V}\right]_- 
= \frac{V_k^*}{\sqrt{L}} \hat{d}_{\sigma}^{\vphantom{+}} &, & 
\left[\hat{d}_{\sigma}^{\vphantom{+}},\hat{V}\right]_- 
= \sum_k \frac{V_k}{\sqrt{L}} \hat{c}_{k,\sigma}^{\vphantom{+}}
\; .
\end{eqnarray}
For non-interacting electrons,
the equations of motion
lead to a closed set of differential equations~(\ref{eq:GFdot1})--(\ref{eq:GFdot4}).

\subsubsection{Fourier transformation of time derivatives}

The equation-of-motion method works in the frequency domain.
The Fourier transformation
of the time derivative of retarded Green functions are given by
\begin{eqnarray}
{\rm FT}\left\{{\rm i}\dot{G}_{A,B}^{\rm ret}(t)\right\}(\omega)&=&
\int_{-\infty}^{\infty}{\rm d} t e^{-\eta |t|} e^{{\rm i}\omega t} 
\left({\rm i} \dot{G}_{A,B}(t)\right) \nonumber \\
&=& {\rm i}\biggl[ 
\left. G_{A,B}^{\rm ret}(t)e^{-\eta |t|}e^{{\rm i}\omega t}\right|_{-\infty}^{\infty}
\nonumber \\
&& - \int_0^{\infty} {\rm d}t G_{A,B}^{\rm ret}(t)
\frac{{\rm d}}{{\rm d}t} 
\left( e^{-\eta t}e^{{\rm i}\omega t}\right)
\biggr] \nonumber \\
&=& (\omega+{\rm i}\eta) \int_{-\infty}^{\infty} {\rm d}t G_{A,B}^{\rm ret}(t)
e^{-\eta |t|}e^{{\rm i}\omega t}\nonumber \\
&=& (\omega+{\rm i}\eta) \tilde{G}_{A,B}^{\rm ret}(\omega) \; ,
\label{eq:FTGdot}
\end{eqnarray}
where we used partial integration in the first step and
the fact that $G_{A,B}^{\rm ret}(t<0)=0$.

The Fourier transformation of a
Green function's time derivative can also be done using contour
integration. 
By definition of the Fourier transformation, we have
\begin{equation}
{\rm i}\dot{G}_{A,B}^{\rm ret}(t)=
\int_{-\infty}^{\infty}\frac{{\rm d} \lambda}{2\pi} 
e^{-\eta_2 |\lambda|} e^{-{\rm i}\lambda t} 
\lambda \tilde{G}_{A,B}^{\rm ret}(\lambda) \; .
\end{equation}
To find the Fourier transformation of the left-hand side
we multiply both sides with ${\rm exp}(-\eta t+{\rm i}\omega t) $
and integrate over $t$ from zero to infinity.
Thus,
\begin{eqnarray}
{\rm FT}\Bigl\{{\rm i}\dot{G}_{A,B}^{\rm ret}(t)\Bigr\}(\omega)&=&
\int_{-\infty}^{\infty}\frac{{\rm d} \lambda}{2\pi} 
e^{-\eta_2 |\lambda|} 
\lambda \tilde{G}_{A,B}^{\rm ret}(\lambda)
\nonumber \\
&& \times  
\int_0^{\infty} {\rm d}t e^{-{\rm i}\lambda t} e^{{\rm i}\omega t-\eta t} \nonumber \\
&=& 
\int_{-\infty}^{\infty}\frac{{\rm d} \lambda}{2\pi{\rm i}} 
\tilde{G}_{A,B}^{\rm ret}(\lambda) e^{-\eta_2 |\lambda|} 
\frac{\lambda}{\lambda-\omega-{\rm i}\eta}  \nonumber \\
&=& (\omega+{\rm i}\eta)
\int_{-\infty}^{\infty}\frac{{\rm d} \lambda}{2\pi{\rm i}} 
\frac{\tilde{G}_{A,B}^{\rm ret}(\lambda) 
e^{-\eta_2 |\lambda|} }{\lambda-\omega-{\rm i}\eta}
\; ,\nonumber \\
\label{eq:FTrealaxis}
\end{eqnarray}
where we used the fact that 
$G^{\rm ret}_{A,B}(t=0^-)=0$ in the last step. Now that 
$\tilde{G}_{A,B}^{\rm ret}(\lambda)$ 
has only poles in the lower half of the complex
plane, we extend the integral over the real axis in equation~(\ref{eq:FTrealaxis})
to a contour integral with an arc of infinite radius
in the upper half of the complex plane.
Since $G_{A,B}^{\rm ret}(\omega)\sim 1/\omega$ as seen from the Lehmann
representation, the arc does not give a finite contribution.
Then, the integral can be evaluated using the residue theorem.
The pole of strength unity at $\lambda=\omega+{\rm i}\eta$
gives, letting $\eta_2=0$, $\eta=0$ where appropriate,
\begin{equation}
{\rm FT}\left\{{\rm i}\dot{G}_{A,B}^{\rm ret}(t)\right\}(\omega)
= (\omega+{\rm i}\eta) \tilde{G}_{A,B}^{\rm ret}(\omega) \; ,
\end{equation}
and we recover equation~(\ref{eq:FTGdot}).

\subsubsection{Explicit solution in the frequency domain}

To solve the equations~(\ref{eq:GFdot1})--(\ref{eq:GFdot4})
we transformation them into frequency space.
We find
\begin{eqnarray}
\left(\omega+{\rm i}\eta\right)
 \tilde{G}_{k,p}^{\rm ret}(\omega)
&=& \delta_{k,p}
+ \epsilon(k) \tilde{G}_{k,p}^{\rm ret}(\omega) 
+\frac{V_k^*}{\sqrt{L}} \tilde{G}_{d,p}^{\rm ret}(\omega)  \; ,\nonumber \\
\label{eq:GFdot1om}\\
\left(\omega+{\rm i}\eta\right)\tilde{G}_{d,p}^{\rm ret}(\omega)&=& 
\sum_k\frac{V_k}{\sqrt{L}} \tilde{G}_{k,p}^{\rm ret}(\omega) \; ,
\label{eq:GFdot2om}\\
\left(\omega+{\rm i}\eta\right)\tilde{G}_{k,d}^{\rm ret}(\omega)&=& 
\epsilon(k) \tilde{G}_{k,d}^{\rm ret}(\omega)
+\frac{V_k^*}{\sqrt{L}} \tilde{G}_{d,d}^{\rm ret}(\omega) \; ,
\label{eq:GFdot3om}\\
\left(\omega+{\rm i}\eta\right)\tilde{G}_{d,d}^{\rm ret}(\omega) &=& 
1 + \sum_k\frac{V_k}{\sqrt{L}} \tilde{G}_{k,d}^{\rm ret}(\omega)\; .
\label{eq:GFdot4om}
\end{eqnarray}
Since we have obtained algebraic equations as a function of $\omega$,
we are now in the position to transform
the retarded to the causal Green function, i.e.,
the equations of motion for the causal Green function 
in frequency space are obtained by 
replacing $\eta$ by $\eta{\rm sgn}(\omega)$ in 
eqs.~(\ref{eq:GFdot1om})--(\ref{eq:GFdot4om}).

The resulting set of equations is readily solved.
We define the retarded and causal hybridization functions
\begin{eqnarray}
\Delta^{\rm ret} (\omega)&=& 
\frac{1}{L} 
\sum_{k}\frac{|V_k|^2}{\omega-\epsilon(k)+{\rm i}\eta}
\nonumber \; ,\\ 
\Delta^{\rm c} (\omega)&=& 
\frac{1}{L} 
\sum_{k}\frac{|V_k|^2}{\omega-\epsilon(k)+{\rm i}\eta{\rm sgn}(\omega)}
\; , 
\end{eqnarray}
and find
\begin{eqnarray}
 \tilde{G}_{k,p}^{\rm c}(\omega) &=& 
\frac{1}{\omega-\epsilon(k)+{\rm i}\eta{\rm sgn}(\omega)}
\biggl( \delta_{k,p}  \nonumber\\
&& +\frac{1}{L}
\frac{V_pV_k^*}{(\omega-\epsilon(p)+{\rm i}\eta{\rm sgn}(\omega))
(\omega-\Delta^{\rm c}(\omega))}
\biggr) \; , \nonumber \\
\label{eq:EOMimpurityGF1}
\end{eqnarray}
\begin{equation}
\tilde{G}_{d,p}^{\rm c}(\omega) = 
\sqrt{\frac{1}{L}}
\frac{V_p}{(\omega-\epsilon(p)+{\rm i}\eta{\rm sgn}(\omega))
(\omega-\Delta^{\rm c}(\omega))}
\; , \label{eq:EOMimpurityGF2}
\end{equation}
\begin{equation}
\tilde{G}_{k,d}^{\rm c}(\omega) =
\sqrt{\frac{1}{L}}
\frac{V_k^*}{(\omega-\epsilon(k)+{\rm i}\eta{\rm sgn}(\omega))
(\omega-\Delta^{\rm c}(\omega))}
\; , \label{eq:EOMimpurityGF3}
\end{equation}
and 
\begin{equation}
\tilde{G}_{d,d}^{\rm c}(\omega) =
\frac{1}{\omega-\Delta^{\rm c}(\omega)}
\; .
\label{eq:EOMimpurityGF4}
\end{equation}
The equations for the retarded Green functions are obtained by replacing
$\eta{\rm sgn}(\omega)$ by $\eta$.

\subsection{Spectral properties}

\subsubsection{Impurity spectral function}

First, we re-derive the impurity spectral function
from the impurity Green function~(\ref{eq:EOMimpurityGF4}).
We have
\begin{eqnarray}
\Delta^{\rm ret}(\omega) &=& \int_{-1/2}^{1/2} {\rm d}\epsilon
\frac{\rho_0(\epsilon)|V(\epsilon)|^2}{\omega-\epsilon+{\rm i}\eta}\nonumber \\
&=&\Lambda_V(\epsilon)-{\rm i}\pi 
\rho_0(\omega)|V(\omega)|^2 \; .
\label{eq:DeltaReandIm}
\end{eqnarray}
The definition of the spectral function immediately gives
\begin{eqnarray}
D_{d,d}(\omega) &=& -\frac{1}{\pi} 
{\rm Im}\left(\frac{1}{\omega-\Delta^{\rm ret}(\omega)}\right)\nonumber \\
&=& \frac{\rho_0(\omega)|V(\omega)|^2}{(\pi\rho_0(\omega)|V(\omega)|^2)^2
+(\omega-\Lambda_V(\omega))^2} \; ,
\end{eqnarray}
as derived in Sect.~\ref{sec:eigenspec}.

\subsubsection{Density of states}

We write the density of states in the form
\begin{eqnarray}
D_{\sigma}(\omega)&=& - \frac{1}{\pi} {\rm Im}\biggl(\sum_m 
\langle \hat{a}_{m,\sigma}^+ \frac{1}{\omega-(\hat{H}_0-E_0)+{\rm i}\eta}
\hat{a}_{m,\sigma}^{\vphantom{+}}\rangle \nonumber \\
&& \hphantom{- \frac{1}{\pi}}
+\langle \hat{a}_{m,\sigma}^{\vphantom{+}}
\frac{1}{\omega-(\hat{H}_0-E_0)+{\rm i}\eta}
\hat{a}_{m,\sigma}^+ 
\rangle
\biggr) \, , 
\end{eqnarray}
where we used the fact that $\hat{a}_{m,\sigma}^+$
($\hat{a}_{m,\sigma}^{\vphantom{+}}$)
creates (annihilates) an electron with energy $E(m)$ in the ground state.
The sum over all $m$ runs over all single-particle
excitations of the ground state and thus represents
the trace over all single-particle eigenstates,
\begin{equation}
D_{\sigma}(\omega)= - \frac{1}{\pi} {\rm Im}{\rm Tr}_1
\Bigl(\frac{1}{\omega-(\hat{H}_0-E_0)+{\rm i}\eta}\Bigr) \; .
\end{equation}
We can equally use the excitations
$\hat{c}_{k,\sigma}^+|\psi_0\rangle$, 
$\hat{c}_{k,\sigma}^{\vphantom{+}}|\psi_0\rangle$,
and 
$\hat{d}_{\sigma}^+|\psi_0\rangle$, 
$\hat{d}_{\sigma}^{\vphantom{+}}|\psi_0\rangle$, respectively,
to perform the
trace over the single-particle excitations of the ground state.
Therefore, we may write
\begin{eqnarray}
D_{\sigma}(\omega)&=& - \frac{1}{\pi} {\rm Im}
\biggl[\sum_k \biggl(
\langle \hat{c}_{k,\sigma}^+ \frac{1}{\omega-(\hat{H}_0-E_0)+{\rm i}\eta}
\hat{c}_{k,\sigma}^{\vphantom{+}}\rangle 
\nonumber \\
&& \hphantom{ - \frac{1}{\pi} {\rm Im}\biggl( \sum_k}
+ \langle \hat{c}_{k,\sigma}^{\vphantom{+}}
\frac{1}{\omega-(\hat{H}_0-E_0)+{\rm i}\eta}
\hat{c}_{k,\sigma}^+ 
\rangle\biggr)
\nonumber \\
&& 
\hphantom{ - \frac{1}{\pi} {\rm Im}\biggl( }
+ 
\langle \hat{d}_{\sigma}^+ \frac{1}{\omega-(\hat{H}_0-E_0)+{\rm i}\eta}
\hat{d}_{\sigma}^{\vphantom{+}}\rangle
\nonumber \\
&& \hphantom{ - \frac{1}{\pi} {\rm Im}\biggl( }
+\langle \hat{d}_{\sigma}^{\vphantom{+}}
\frac{1}{\omega-(\hat{H}_0-E_0)+{\rm i}\eta}
\hat{d}_{\sigma}^+ 
\rangle \biggr] \nonumber \\
&=& - \frac{1}{\pi} {\rm Im}
\biggl[
\sum_k G_{k,k}^{\rm ret}(\omega)+ G_{d,d}^{\rm ret}(\omega) 
\biggr] \; .
\end{eqnarray}
Equation~(\ref{eq:EOMimpurityGF1}) shows that the band Green function
consists of the undisturbed host Green function for $V_k\equiv 0$
and a $1/L$ correction due to the hybridization.
Therefore, using eqs.~(\ref{eq:EOMimpurityGF1}) and~(\ref{eq:EOMimpurityGF4}),
the contribution due to a finite hybridization is given by
\begin{eqnarray}
D_{{\rm imp},\sigma}(\omega)&=& -\frac{1}{\pi}{\rm Im}
\biggl[
\frac{1}{\omega-\Delta^{\rm ret}(\omega)}\nonumber \\
&& \hphantom{-\frac{1}{\pi}{\rm Im}}\times 
\left(
1 + \sum_k 
\frac{|V_k|^2/L}{(\omega-\epsilon(k)+{\rm i}\eta)^2}
\right)
\biggr] \nonumber \\
&=& 
- \frac{1}{\pi} {\rm Im}
\left[
\frac{1-(\partial \Delta^{{\rm ret}}(\omega))/(\partial\omega)}{
\omega-\Delta^{\rm ret}(\omega)}\right]\nonumber \\ 
&=& 
- \frac{1}{\pi} \frac{\partial }{\partial \omega}
{\rm Im}
\left[\ln\left(\omega-\Delta^{\rm ret}(\omega) \right)\right]
\; .
\end{eqnarray}
We use equation~(\ref{eq:DeltaReandIm}) and find from the complex logarithm
\begin{eqnarray}
D_{{\rm imp},\sigma}(\omega)&=& - \frac{1}{\pi} 
\frac{\partial }{\partial \omega}
\left[
\tan^{-1}\left( 
\frac{\pi\rho_0(\omega)|V(\omega)|^2}{\omega-\Lambda_V(\omega)}
\right)
\right]\nonumber \\
&=& 
\frac{\partial }{\partial \omega}
\left[
\rho_0(\omega)y(\omega)\right] \; ,
\label{eq:DOSfromGFagain}
\end{eqnarray}
as derived in Sect.~\ref{sec:eigenspec}.

\subsection{Ground-state expectation values}

Lastly, we re-derive the ground-state expectation values
for the $d$-occupancy, the hybridization matrix element, and
the momentum distribution from the Green function approach.

\subsubsection{Expectation values from Green functions}

The Green functions permit the calculation of ground-state expectation values.
By definition, we have ($\eta=0^+$)
\begin{eqnarray}
\langle \hat{B} \hat{A}\rangle 
&=& (-{\rm i}) G_{A,B}(t=-\eta) \nonumber \\
&=& \int_{-\infty}^{\infty} \frac{{\rm d}\omega}{2\pi{\rm i}}
e^{-\eta_2|\omega|} e^{{\rm i}\eta\omega}
\tilde{G}_{A,B}^{\rm c}(\omega) \; .
\end{eqnarray}
We extend the integral over the real axis into a contour integral
in the complex plane where the closed contour~$C$
runs over the real axis and an arc with infinite radius
in the upper complex plane. Due to the factor 
$\exp[{\rm i}\eta ({\rm Re}(z)+{\rm i}{\rm Im}(z)]$, the arc does 
not contribute because ${\rm Im}(z)\to +\infty$ on the arc.
Therefore, we have
\begin{equation}
\langle \hat{B} \hat{A}\rangle = 
\oint_{C}\frac{{\rm d}z}{2\pi{\rm i}}\tilde{G}_{A,B}^{\rm c}(z) e^{{\rm i}z\eta}\; .
\label{eq:gsfromGF}
\end{equation}
It is not always easy to do the integral because the Green 
functions display branch cuts in the complex plane.

\subsubsection{Ground-state energy}
The ground-state energy can immediately be calculated using the density of states,
\begin{equation}
\frac{\Delta E}{2}=\int_{-1/2}^0{\rm d}\omega \omega D_{{\rm imp},\sigma}(\omega)
\; .
\end{equation}
The result for the impurity density of states~(\ref{eq:DOSfromGFagain})
and a partial integration directly
lead to the desired result for the ground-state energy.

\subsubsection{Impurity occupancy}

For the $d$-electron occupancy,
$\hat{A}=\hat{d}_{\sigma}^{\vphantom{+}}$, 
$\hat{B}=\hat{d}_{\sigma}^+$, we find 
\begin{equation}
\langle \hat{d}_{\sigma}^+ \hat{d}_{\sigma}^{\vphantom{+}}\rangle = 
\oint_{C}\frac{{\rm d}z}{2\pi{\rm i}}e^{{\rm i}\eta z}
\frac{1}{z-\Delta^{\rm c}(z)} \; .
\end{equation}
$\Delta^{\rm c}(z)$ has a branch cut on the real axis 
that is infinitesimally above the real axis for ${\rm Re}(z)=\omega<0$
and infinitesimally below the real axis for $\omega>0$.
Since $1/(z-\Delta^{\rm c}(z))$ is otherwise
analytic in the complex plane, we can deform the contour~$C$
to $\tilde{C}$ where $\tilde{C}$ encloses the branch cut
for $-1/2<\omega<0$ at infinitesimal distance $\xi$. 
The corners of $\tilde{C}$ provide a vanishingly
small contribution and only the integrals below and above the branch cut
remain finite for $\xi\to 0$,
\begin{eqnarray}
\langle \hat{d}_{\sigma}^+ \hat{d}_{\sigma}^{\vphantom{+}}\rangle &=& 
\oint_{\tilde{C}}\frac{{\rm d}z}{2\pi{\rm i}}
\frac{1}{z-\Delta^{\rm c}(z)} \nonumber \\
&=& 
\int_{-1/2}^0 \frac{{\rm d}\omega}{2\pi {\rm i}}
\frac{1}{\omega-\Lambda_V(\omega)-{\rm i}\rho_0(\omega)|V(\omega)|^2}
\nonumber \\
&& +
\int_{0}^{-1/2} \frac{{\rm d}\omega}{2\pi {\rm i}}
\frac{1}{\omega-\Lambda_V(\omega)+{\rm i}\rho_0(\omega)|V(\omega)|^2}
 \; , \nonumber \\
\label{eq:GFdoccupancy}
\end{eqnarray}
where we used 
\begin{eqnarray}
\Delta^{\rm c}(\omega-{\rm i}\xi)&=&
\Lambda_V(\omega)+{\rm i}\pi \rho_0(\omega)|V(\omega)|^2
\; ,\nonumber \\
\Delta^{\rm c}(\omega+{\rm i}\xi)&=&
\Lambda_V(\omega)-{\rm i}\pi \rho_0(\omega)|V(\omega)|^2 
\end{eqnarray}
infinitesimally below and above the branch cut. From equation~(\ref{eq:GFdoccupancy})
we readily recover 
$\langle \hat{d}_{\sigma}^+ \hat{d}_{\sigma}^{\vphantom{+}}\rangle =1/2$.

\subsubsection{Hybridization}

The derivation of the hybridization matrix element
proceeds along the same lines. We have
\begin{equation}
\langle \hat{c}_{k,\sigma}^+ \hat{d}_{\sigma}^{\vphantom{+}}\rangle = 
\oint_{C}\frac{{\rm d}z}{2\pi{\rm i}}
\frac{1}{z-\epsilon(k)+{\rm i}{\rm sgn}(\epsilon(k))}
\frac{1}{z-\Delta^{\rm c}(z)} \; .
\end{equation}
For $0<\epsilon(k)<1/2$, there is a pole in the lower complex plane
that does not give a contribution to the contour integral.
Following the same lines as for the impurity occupancy we thus find
\begin{eqnarray}
\langle \hat{c}_{k,\sigma}^+ \hat{d}_{\sigma}^{\vphantom{+}}\rangle &=& 
 \frac{V_k}{\sqrt{L}}\int_{-1/2}^0
{\rm d}\omega
\frac{1}{\omega-\epsilon(k)}\nonumber \\
&& \hphantom{ \frac{V_k}{\sqrt{L}}}
\times \frac{\rho_0(\omega)|V(\omega)|^2}{
[\omega-\Lambda_V(\omega)]^2+[\pi \rho_0(\omega)|V(\omega)|^2]^2}
\nonumber \\
&=& \frac{V_k}{\sqrt{L}} H(\epsilon(k))
\end{eqnarray}
for $\epsilon(k) >0$ with $H(\epsilon)$ from the main text.
This contribution is also present for $-1/2<\epsilon(k)<0$ but
the integral must be understood as principal value integral to circumvent
the singularity at $\omega=\epsilon(k)$.
For $-1/2<\epsilon(k)<0$, our contour $\tilde{C}$ also encloses the
pole at $z=\epsilon+{\rm i}\eta$. The pole contributes at the real
value $\omega=\epsilon(k)$, i.e., on the branch cut itself where
\begin{equation}
{\rm Re}\left(\frac{1}{\omega-\Delta^{\rm c}(\omega)}\right)
= \frac{\omega-\Lambda_V(\omega)}{
[\omega-\Lambda_V(\omega)]^2+[\pi \rho_0(\omega)|V(\omega)|^2]^2}\; .
\end{equation}
Thus, we find 
\begin{eqnarray}
\langle \hat{c}_{k,\sigma}^+ \hat{d}_{\sigma}^{\vphantom{+}}\rangle &=& 
\frac{V_k}{\sqrt{L}}H(\epsilon(k)) \nonumber \\
&& +
\frac{(V_k/\sqrt{L})\rho_0(\epsilon(k))|V(\epsilon(k))|^2}{
[\epsilon(k)-\Lambda_V(\epsilon(k))]^2+
[\pi \rho_0(\epsilon(k))|V(\epsilon(k))|^2]^2}\nonumber \\
&=& \frac{V_k}{\sqrt{L}} \left[H(\epsilon(k)) +G(\epsilon(k))\right]
\end{eqnarray}
for $\epsilon(k) <0$ with $G(\epsilon)$ from Sect.~\ref{sec:gsexpectationvalues}.

\subsubsection{Momentum distribution}

The calculation of the momentum distribution 
$n_{k,\sigma}=n_{k,\sigma}^{(0)}+n_{k,\sigma}^{(1)}/L$
requires the elementary integral
\begin{equation}
n_{k,\sigma}^{(0)}
= 
\oint_{C}\frac{{\rm d}z}{2\pi{\rm i}}e^{{\rm i}\eta z}
\frac{1}{z-\epsilon(k)+{\rm i}{\rm sgn}(\epsilon(k))}
= \Theta(-\epsilon(k)) \; .
\end{equation}
Here, we used the fact that there is a pole in the upper complex plane 
of strength unity for $\epsilon(k)<0$ only.
Moreover,
\begin{eqnarray}
n_{k,\sigma}^{(1)} &=& |V_k|^2
\int_{-\infty}^{\infty} \frac{{\rm d}\omega}{2\pi{\rm i}}
e^{{\rm i}\eta\omega} \frac{1}{\omega-\Delta^{\rm c}(\omega)}\nonumber \\
&& \hphantom{ |V_k|^2\int_{-\infty}^{\infty}}
\times
\frac{1}{[\omega-\epsilon(k)+{\rm i}\eta{\rm sgn}(\omega)]^2}
\nonumber \\
&=& |V_k|^2\frac{\partial}{\partial \epsilon(k)}
\left[G(\epsilon(k))+H(\epsilon(k))\right]\nonumber\\
&=& 
|V_k|^2\left[G'(\epsilon(k))+H'(\epsilon(k))\right]
 \; ,
\end{eqnarray}
as derived in Sect.~\ref{sec:gsexpectationvalues}.

\begin{acknowledgements}
We thank Stefan Kehrein for bringing appendix~I of
Ref.~\cite{Schoeller1998} to our attention.
Z.M.M.\ Mahmoud thanks his colleagues at the
Department of Physics in Marburg for their hospitality.
\end{acknowledgements}

\providecommand{\WileyBibTextsc}{}
\let\textsc\WileyBibTextsc
\providecommand{\othercit}{}
\providecommand{\jr}[1]{#1}
\providecommand{\etal}{~et~al.}


\begin{thebibliography}{[1]}

\bibitem{Anderson1961}
 \textsc{P.\,W. Anderson},
 \jr{Phys. Rev.} \textbf{124}, 41 (1961).


\othercit
\bibitem{Hewsonbook}
 \textsc{A.\,C. Hewson},
{The} {Kondo} {Problem} to {Heavy} {Fer\-mions}, Cambridge Studies in
  Magnetism,  Vol.\,2 (Cambridge University Press, Cambridge, 1997).


\othercit
\bibitem{Solyombook}
 \textsc{J.~S\'{o}lyom}, {Fundamentals} of the {Physics} of {Solids},
 Vol.\,3 (Springer, Heidelberg, 2010).


\bibitem{Schoenhammer1990}
 \textsc{K.~Sch\"onhammer},
 \jr{Phys. Rev. B} \textbf{42}, 2591 (1990).


\bibitem{Gebhard1991}
 \textsc{F.~Gebhard},
 \jr{Phys. Rev. B} \textbf{44}, 992 (1991).


\othercit
\bibitem{GR}
 \textsc{I.\,S. Gradshteyn} and  \textsc{I.\,M. Ryzhik},
{Table} of {Integrals}, {Series} and {Products}, eighth edition, ed.\ by D.
  Zwillinger and V. Moll (Academic Press, Amsterdam, 2015).


\bibitem{Schoeller1998}
 \textsc{J.~von Delft} and  \textsc{H.~Schoeller},
 \jr{Ann. Phys.} \textbf{7}, 225--305 (1998).


\end{thebibliography}

\end{document}